\documentclass[pre,twocolumn,twoside,showpacs,superscriptaddress,floatfix]{revtex4-1}
\usepackage{graphics,amssymb,amsmath,epsf,color,hyperref}
\epsfclipon
\usepackage{epsfig,bm,epsf,graphics}
\usepackage{graphicx}
\usepackage{dcolumn}
\usepackage{bm}
\usepackage{braket}
\usepackage{color}
\usepackage{soul}
\usepackage{xcolor}

\newcommand{\jo}{\textcolor{black}}

\begin{document}

\title{{Data augmentation using diffusion models to enhance inverse Ising inference}}

\author{Yechan Lim}
\affiliation{Department of Physics Education, Seoul National University, Seoul 08826, Korea}

\author{Sangwon Lee}
\affiliation{{Department of Physics, The University of Tokyo, Tokyo 113-0033, Japan}}

\author{Junghyo Jo}
\email{jojunghyo@snu.ac.kr}
\affiliation{Department of Physics Education and Center for Theoretical Physics and Artificial Intelligence Institute, Seoul National University, Seoul 08826, Korea}
\affiliation{School of Computational Sciences, Korea Institute for Advanced Study, Seoul 02455, Korea}

\date{\today}

\begin{abstract}
Identifying model parameters from observed configurations poses a fundamental challenge in data science, especially with limited data. Recently, diffusion models have emerged as a novel paradigm in generative machine learning, capable of producing new samples that closely mimic observed data. These models learn the gradient of model probabilities, bypassing the need for cumbersome calculations of partition functions across all possible configurations. We explore whether diffusion models can enhance parameter inference by augmenting small datasets. Our findings demonstrate this potential through a synthetic task involving inverse Ising inference and a real-world application of reconstructing missing values in neural activity data. This study serves as a proof-of-concept for using diffusion models for data augmentation in physics-related problems, thereby opening new avenues in data science.

\end{abstract}

\maketitle


\section{Introduction}

Physics has significantly influenced the development of machine learning~\cite{Karniadakis2021Jun}. One notable example is the Hopfield model, a spin glass model in physics, which serves as the prototype for recent deep neural networks~\cite{Hopfield1982Apr}. On a more theoretical side, renormalization group approaches provide insights into how deep neural networks operate as information transfer processes~\cite{Mehta2014Oct,Koch-Janusz2018Jun}. Other fields of study in physics, such as stochastic thermodynamics, further help scientists understand what happens during the training of neural networks and how to utilize this knowledge to even improve current machine learning algorithms.~\cite{Goldt2017Jan}

Conversely, machine learning has made substantial contributions to physics~\cite{Carleo2019Dec}. For instance, supervised learning techniques can identify phase transitions in statistical mechanics and decipher collective behaviors of active matter~\cite{Carrasquilla2017May,Cichos2020Feb}. Furthermore, unsupervised learning techniques such as Boltzmann machine can generate unseen samples~\cite{Ackley1985Jan}, while autoencoders can reduce the dimensionality of input data, creating coarse-grained latent representations and denoising input data~\cite{Vincent2008Jul}. Variational autoencoders~\cite{Kingma2013Dec} \jo{and generative adversarial networks~\cite{goodfellow2014generative}} play a role by generating new samples that mimic observed data. {These various machine learning techniques allow physicists to extract valuable information from massive and noisy datasets.}

Recently, diffusion models have emerged, drawing on concepts from physics to advance their design and application~\cite{sohldickstein2015deep}. Unlike traditional models that represent scalar functions of model probabilities, diffusion models focus on learning the gradient of these probabilities in response to small perturbations in data configurations. In the framework of stochastic differential equations, this gradient vector is referred to as the ``score" function, analogous to the force field in molecular dynamics. Consequently, the evolution of model probabilities is governed by a Langevin equation incorporating these score vectors, rather than the Fokker-Planck equation, which describes the evolution of the probability distribution itself. These diffusion models~\cite{sohldickstein2015deep,ho2020denoising,song2020improved} have demonstrated remarkable success in generating high-quality outputs across diverse domains, including images, videos, and audio~\cite{kong2021diffwave}, natural language~\cite{austin2023structured}, temporal data~\cite{tashiro2021csdi}, and drug molecules design~\cite{guan20233d}, establishing themselves as some of the most robust generative models in machine learning today~\cite{Yang2023Nov}.


In this study, we aim to explore whether the generative capabilities of diffusion models can be utilized to address physics problems. Specifically, we investigate the potential of these models to augment small datasets and enhance the inference of model parameters.

As a concrete illustration, consider a binary pattern $\sigma =(\sigma_{1}, \sigma_{2},\cdots,\sigma_{n})$ in $n$ dimensions, where $\sigma_{i} = \pm1$. This pattern follows a Boltzmann distribution:
\begin{equation}
\label{eq:Boltzmann weight}
p(\sigma)=\frac{\exp(-E(\sigma))}{Z},
\end{equation}
where the partition function $Z=\sum_{\sigma} \exp(-E(\sigma))$ ensures the normalization $\sum_\sigma p(\sigma)=1$.
The energy function is defined as
\begin{equation}
\label{eq:hamiltonian}
E(\sigma) = -\sum_{i}  b_{i}\sigma_{i} -\sum_{i<j}w_{ij}\sigma_{i}\sigma_{j},
\end{equation}
where the bias parameter $b_i$ controls the activity of $\sigma_i$, and the coupling parameter $w_{ij}$ controls the correlation between $\sigma_i$ and $\sigma_j$.
In physics, given $b_i$ and $w_{ij}$, deriving observable quantities and statistical properties such as the magnetization, correlation, and phase transitions from microscopic laws governing the constituents of a system holds significant importance.
In data science, however, the pursuit of identifying model parameters from given configurations stands as a paramount objective. Given a set of $M$ patterns $\{\hat{\sigma}(\mu)\}_{\mu=1}^M$, inferring the model parameters $b_i$ and $w_{ij}$ is an important problem known as {\it inverse Ising inference}~\cite{Nguyen2017Feb}.
Note that we use the notation $\hat{\sigma}$ to represent observed patterns for clarification.

The inverse Ising inference has broad applications for understanding protein structures~\cite{protein,Cocco_2018}, atomic interactions that result in specific crystal lattices~\cite{PhysRevE.88.042309}, financial market dynamics~\cite{Zhao_2018, borysov_2015}, gene recombination processes~\cite{mora_2010}, and patterns of human interaction~\cite{friendship}. Thus, it is a critical tool in many scientific domains. 
However, if the amount of observed data is too small, which is mostly the case when it comes to inverse Ising problem, the inference suffers from noise and the parameters are inferred inaccurately.

In this study, we investigate the use of diffusion models for data augmentation. This paper is organized as follows: In Sec.~\ref{method}, we introduce the foundational tools for inverse Ising inference and describe the application of diffusion models for generating augmented samples. Section~\ref{results} presents our findings, demonstrating the proof-of-concept that augmenting data with diffusion models enhances the quality of inverse Ising inference on both synthetic and real-world datasets. Finally, in Sec.~\ref{discussion}, we summarize our results and discuss potential applications and limitations of this approach.
\jo{Computational details are provided in Appendix~\ref{sec:appendix}.}

\section{Method}
\label{method}
We begin with introducing maximum likelihood estimation~\cite{Nguyen2017Feb} as a standard method and the erasure machine~\cite{PhysRevE.101.032107} as a practical approach for inverse Ising inference. Next, we discuss diffusion models and their role in data generation.

\subsection{Maximum likelihood estimation}
Given data $\{\hat{\sigma}(\mu)\}_{\mu=1}^M$, we can define the data distribution as
\begin{equation}
    f(\sigma) = \frac{1}{M} \sum_{\mu=1}^M \delta(\sigma - \hat{\sigma}(\mu))
\end{equation}
by simply counting the relative frequency of $\sigma$.
We can then define the difference between the data distribution $f(\sigma)$ and the model distribution $p(\sigma)$ using the Kullback-Leibler divergence:
\begin{align}
    D_{KL}(f||p) &\equiv \sum_\sigma f(\sigma) \log \frac{f(\sigma)}{p(\sigma)} \nonumber \\
    &= \sum_\sigma f(\sigma) \log f(\sigma) - \sum_\sigma f(\sigma) \log p(\sigma).
\end{align}
To optimize the model parameters, we minimize $D_{KL}(f||p)$.
Since the first term does not include the model parameters, the optimization is equivalent to maximizing the log-likelihood of data:
\begin{align}
    \mathcal{L} &\equiv \frac{1}{M} \log \prod_{\mu=1}^M p(\hat{\sigma}(\mu)) \nonumber \\
    &= \sum_\sigma f(\sigma) \log p(\sigma) \nonumber \\
    &= \sum_\sigma E(\sigma) f(\sigma) - \log Z.
\end{align}
We can optimize the model parameters using gradient ascent method:
\begin{align}
    b_i &\leftarrow b_i + \gamma \frac{\partial \mathcal{L}}{\partial b_i}, \\
    w_{ij} &\leftarrow w_{ij} + \gamma \frac{\partial \mathcal{L}}{\partial w_{ij}}, 
\end{align}
where $\gamma$ is the learning rate (set to $\gamma=0.1$ for this study).
The gradients can be calculated as follows:
\begin{align}
    \frac{\partial \mathcal{L}}{\partial b_i} &= \sum_\sigma \sigma_i f(\sigma) - \sum_\sigma \sigma_i p(\sigma).
\end{align}
The first and second terms correspond to the data and model averages of $\sigma_i$, respectively.
Similarly, we obtain
\begin{align}
    \frac{\partial \mathcal{L}}{\partial w_{ij}} &= \sum_\sigma \sigma_i \sigma_j f(\sigma) - \sum_\sigma \sigma_i \sigma_j p(\sigma),
\end{align}
which corresponds to the difference in averaged correlations between the data and model distributions.

\subsection{Erasure machine}
The average of observables $O = \{\sigma_i, \sigma_{i}\sigma_{j}\}$ for data and model distributions can be represented as follows:
\begin{align}
    \langle O \rangle_f &\equiv \sum_\sigma O(\sigma) f(\sigma) = \sum_{\hat{\sigma}} O(\hat{\sigma}) f(\hat{\sigma}), \\
    \langle O \rangle_p &\equiv \sum_\sigma O(\sigma) p(\sigma),
\end{align}
where $\hat{\sigma}$ denotes observed configurations among all possible ones, as defined previously. Since $f(\sigma)$ is zero for unobserved configurations $\sigma$, the first equation is valid. This implies that the summation in the first equation is computationally inexpensive as it involves only observed configurations, whereas the summation in the second equation is computationally expensive as it involves all configurations.
For example, for $n=20$ dimensions, $2^{20} \approx 1,000,000$ binary configurations should be considered.
Therefore, maximum likelihood estimation based on the model averages becomes impractical for large systems.

Addressing the computational intractability has been a significant hurdle for physicists. Various approximate techniques have been developed, including mean-field theories~\cite{PhysRevE.58.2302,Ricci-Tersenghi_2012,Nguyen2012Aug}, optimization functions such as pseudolikelihood~\cite{Aurell2012Mar,Decelle2014Feb,Lokhov2018Mar}, and alternative approaches such as minimum probability flow~\cite{Sohl-Dickstein2011Nov} and adaptive cluster expansion~\cite{Cocco2011Mar}. State-of-the-art machine learning techniques such as variational autoregressive networks have also been employed to solve the inverse Ising inference~\cite{PhysRevLett.122.080602}. Notably, Jo \textit{et al.} proposed the \textit{erasure machine} which offers fast and precise inference for both synthetic and real-world data~\cite{PhysRevE.101.032107}.
The idea is simple that instead of direct comparing $f$ and $p$, the erasure machine compares reweighted distributions $f_\epsilon(\sigma)$ and $p_\epsilon (\sigma)$ which correspond to high-temperature distributions in statistical mechanics:
\begin{align}
\label{eq:reweight}
    f_\epsilon(\sigma) &= \frac{f(\sigma) p^{-1+\epsilon}(\sigma)}{\sum_{\sigma'} f(\sigma') p^{-1+\epsilon}(\sigma')} \\
    p_\epsilon(\sigma) &= \frac{p(\sigma) p^{-1+\epsilon}(\sigma)}{\sum_{\sigma'} p(\sigma') p^{-1+\epsilon}(\sigma')} = \frac{p^{\epsilon}(\sigma)}{Z_\epsilon},
\end{align}
where $Z_\epsilon \equiv \sum_{\sigma'} p(\sigma') p^{-1+\epsilon}(\sigma')$.
The hyperparameter $\epsilon$ acts as the inverse temperature $\beta$ in statistical mechanics.
Then, unlike the minimization of $D_{KL}(f||p)$ in the maximum likelihood estimation, the erasure machine minimizes $D_{KL}(f_\epsilon||p_\epsilon)$.
The data average of observables in the high-temperature limit can be calculated as previously:
\begin{align}
    \langle \sigma_i \rangle_{f_\epsilon} &= \sum_{\hat{\sigma}} \hat{\sigma}_i f_\epsilon(\hat{\sigma}), \\
    \langle \sigma_i \sigma_j \rangle_{f_\epsilon} &= \sum_{\hat{\sigma}} \hat{\sigma}_i \hat{\sigma}_j f_\epsilon(\hat{\sigma}).
\end{align}
Interestingly, the model average in the high-temperature limit can be simplified as:
\begin{align}
    \langle \sigma_i \rangle_{p_\epsilon} &\approx \epsilon b_i, \\
    \langle \sigma_i \sigma_j \rangle_{p_\epsilon} &\approx \epsilon w_{ij}.
\end{align}
Notably no summation over all configurations is required in the high-temperature expansion. 
The process involves the following steps:
First, obtain the reweighted data distribution $f_\epsilon(\sigma)$ using the current model distribution $p(\sigma)$ as in Eq.~(\ref{eq:reweight}). 
Second, update model parameters as follows:
\begin{align}
\label{eq:EM update}
b_i &\leftarrow b_{i}+\gamma \bigg(\langle \sigma_{i}\rangle_{f_{\epsilon}}-\epsilon b_{i} \bigg),\nonumber \\
w_{ij} &\leftarrow w_{ij}+\gamma \bigg(\langle \sigma_{i} \sigma_j \rangle_{f_{\epsilon}}-\epsilon w_{ij} \bigg).
\end{align}
Given these parameter updates, revise the model distribution $p(\sigma)$ and repeat the procedure.
The iteration stops when the average energy $\langle E(\sigma) \rangle_f$ is minimized. The energy minimization corresponds to maximizing the probability $p(\hat{\sigma})$ of the observed data pattern $\hat{\sigma}$.
For fine-tuning the hyperparameter $\epsilon$, refer to the original work~\cite{PhysRevE.101.032107}.

\begin{figure}[t]
\centering
\includegraphics[width=7.5cm]{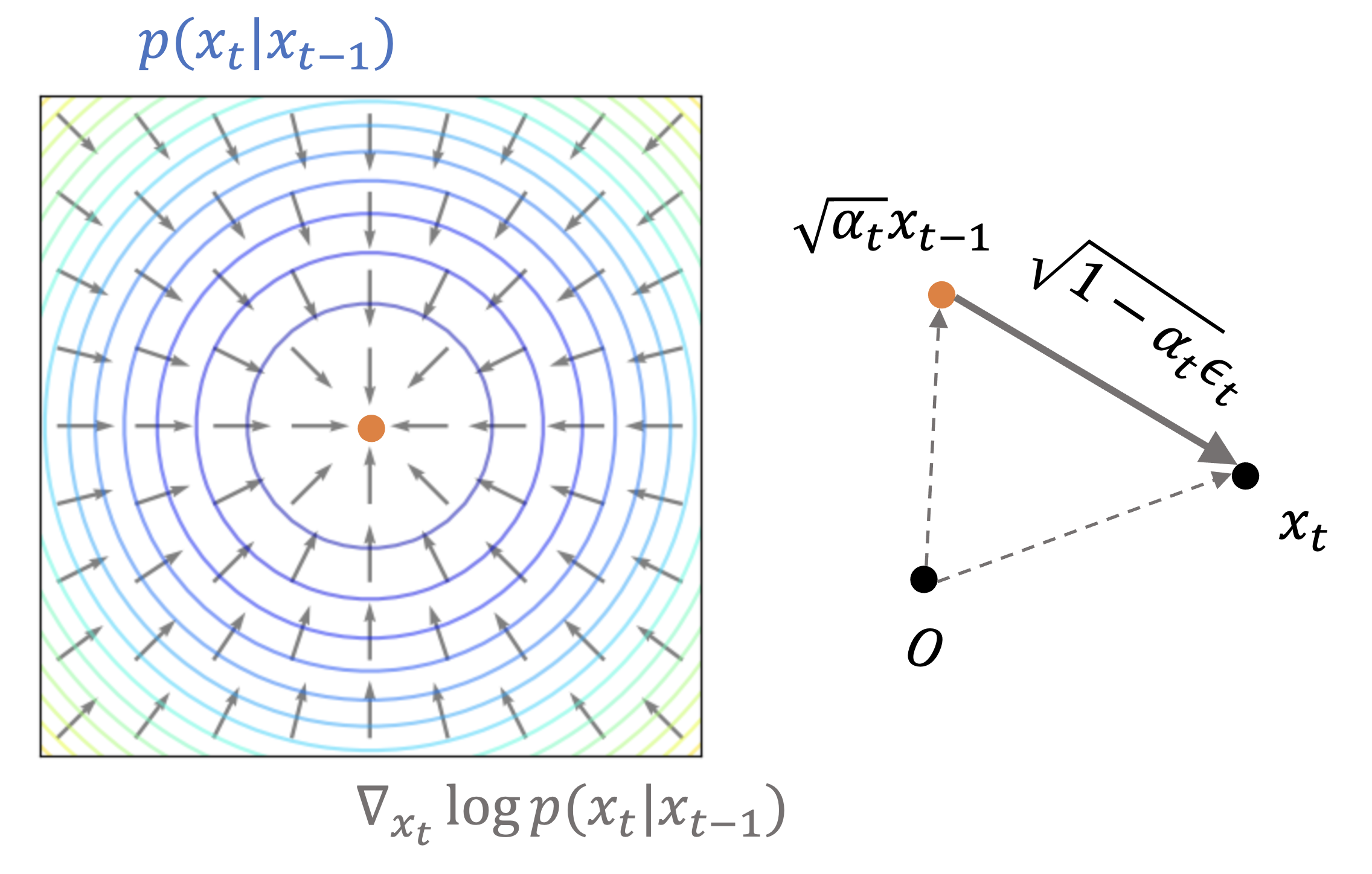}
\caption{\label{fig:diffusion}(Color Online) Schematic diagram of diffusion models. Diffusion models represent a scalar probability function (colored contours) in terms of vector flows (gray arrows). The direction of the vector flows is opposite to the noise vector $\epsilon_t$, which transports the previous $x_{t-1}$ to the present $x_t$.
Noise schedule $\alpha_t$ controls the degree of diffusion.}
\end{figure}

\subsection{Diffusion models}

Suppose we have a sample $x_0$ of data.
We then add noise following a diffusion process:
\begin{equation}
\label{eq:forward}
    x_t = \sqrt{\alpha_t} x_ {t-1}+ \sqrt{1-\alpha_t} \epsilon_t,
\end{equation}
where $\epsilon_t \sim \mathcal{N}(0, I)$ is a Gaussian noise, and $\alpha_t$ is a pre-designed noise schedule.
If we repeat the forward process, $x_0 \to x_1 \to \cdots \to x_T$, finally $x_0$ transports into $x_T$ which follows a Gaussian distribution.
\jo{We use a total of $T=1,000$ transformation steps.}
The Langevin equation of Eq.~(\ref{eq:forward}) can also be represented by a Markov process with a transition probability:
\begin{equation}
    p(x_t|x_{t-1}) = \frac{1}{Z} \exp\bigg[-\frac{(x_t - \sqrt{\alpha_t}x_{t-1})^2}{2(1-\alpha_t)} \bigg].
\end{equation} 
As illustraded in Fig.~\ref{fig:diffusion}, the information of the probability landscape of $p(x_t|x_{t-1})$ can be equally represented by its log gradient:
\begin{equation}
    s(x_t|x_{t-1}) \equiv \nabla_{x_t} \log p(x_t|x_{t-1}),
\end{equation}
which is called the ``score function''~\cite{song2020improved}. 
It is noteworthy that, unlike the scalar function of the probability, the vector function of the score does not require the explicit computation of the partition function $Z$ which can often be cumbersome in problem solving.
Since $p(x_t|x_{t-1})$ is a Gaussian distribution, it is straightforward to show the relation between the score function and the noise~\cite{ho2020denoising}:
\begin{equation}
    \epsilon_t = - \sqrt{1-\alpha_t} s(x_t | x_{t-1}).
\end{equation}
The noise vector indicates the direction of $x_{t-1} \to x_{t}$, whereas the score vector indicates the gradient of the log probability, which points in the direction opposite to the noise vector.
Here, the mapping from $x_t$ to $\epsilon_t$ is realized by neural networks such as the U-Net~\cite{ho2020denoising}: $\epsilon_t =\text{NN}(x_t, t)$.
\jo{The detailed computational implementation of the diffusion model is provided in Appendix~\ref{sec:appendix}.}

By learning the score function or the noise vector from the forward process, we can reverse the diffusion process. The backward process can then be used to generate new data $x_0 \leftarrow x_1 \leftarrow \cdots \leftarrow x_{T-1} \leftarrow x_T$, starting from completely noisy samples $x_T$. Note that the new samples $x_0$ are not part of the training samples $x_0$. To advance this idea to practical stages, the denoising diffusion probabilistic model (DDPM)~\cite{ho2020denoising} and the denoising diffusion implicit model (DDIM)~\cite{song2022denoising} play crucial roles.

In our study, we have binary data $\{\hat{\sigma}(\mu)\}_{\mu=1}^M$. 
A sample $\hat{\sigma}(\mu)$ corresponds to an original data $x_0$ in the diffusion models.
Although a diffusion model for binary images exists~\cite{sohldickstein2015deep}, we found that the usual diffusion models for continuous images work better for our task. 
Thus we train diffusion models with $\{x_0(\mu)\}_{\mu=1}^M = \{\hat{\sigma}(\mu)\}_{\mu=1}^M$. 
Using the backward process, we generate new $x_0$ samples. To obtain binary patterns of $\sigma$, we use analogue bits method, which decodes $x_0$ into binary $\sigma$ through a thresholding operation~\cite{chen2023analog}.


\begin{figure*}[th]
\centering
\includegraphics[width=0.85\textwidth]{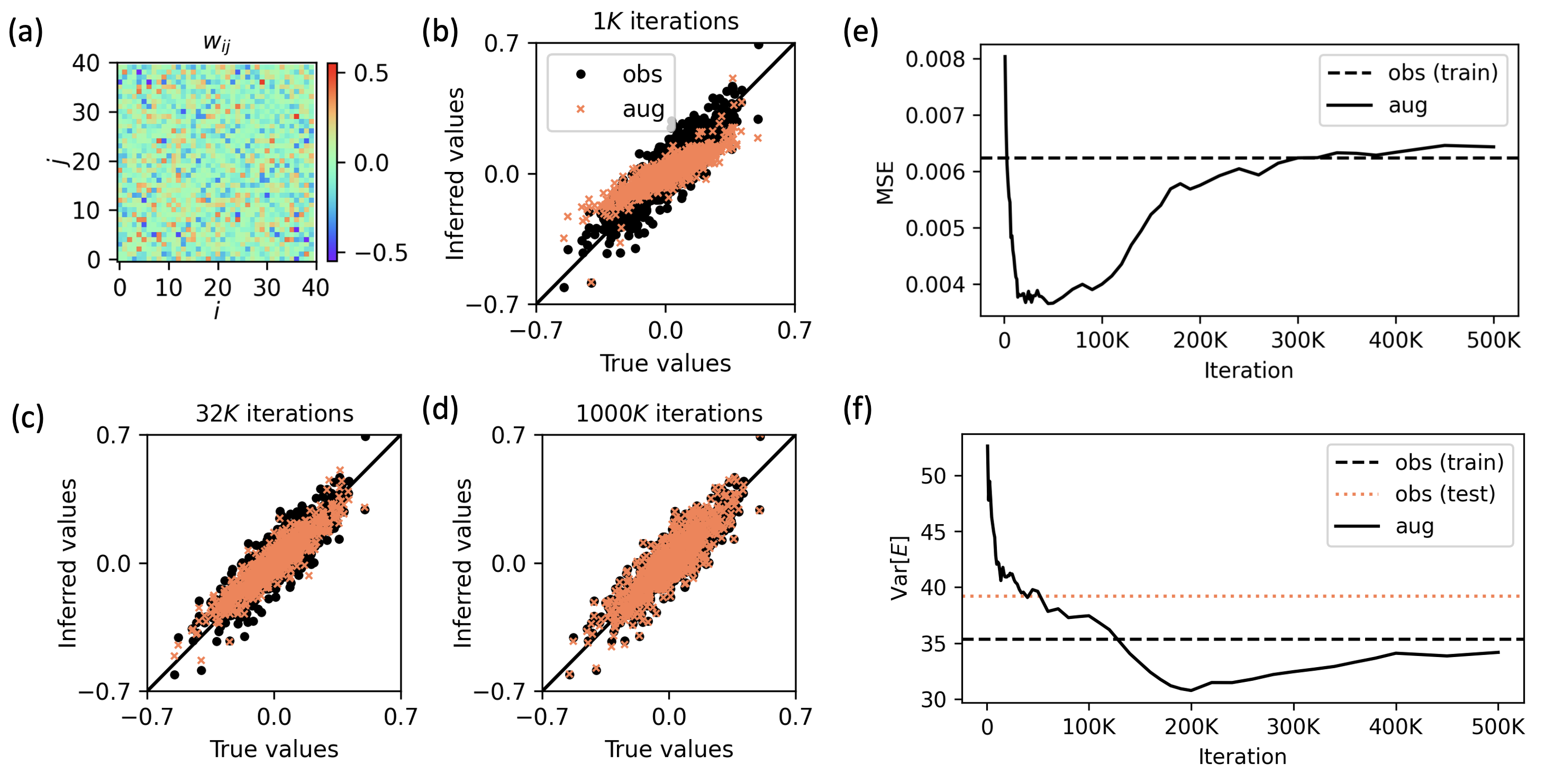}
\caption{\label{iteration plot} (Color Online)
Inverse Ising inference with augmented data. 
(a) Heatmap of interaction parameters, sampled from a normal distribution $\mathcal{N}(0, g^{2}/n)$ with $g=1$, in the Sherrington-Kirkpatrick model in dimensions $n=40$. 
Inferred parameter values versus true parameter values are compared for the inference results using only $M=4,000$ observed data (filled black circles) and using $M^+=100,000$ augmented data (orange crosses) generated by the diffusion model after (b) 1,000, (c) 32,000, and (d) 1,000,000 iterations. 
(e) The inference performance, measured by the mean square error (MSE) between inferred and true parameter values, varies depending on the degree of learning of the diffusion model. The MSE from the inference using only observed data is shown as a reference by the dashed line. 
(f) Variances of the energies of observed binary patterns in the training and test sets, as well as the energies of the augmented data, are measured and compared at different learning stages of the diffusion model, with all energies measured using the  $\{b_i, w_{ij}\}$ values inferred solely from the training data. Dashed and dotted lines represent the reference energy variances for the observed data in the training and test sets, respectively.}
\end{figure*}

\begin{figure}[t]
\centering
\includegraphics[width=0.4\textwidth]{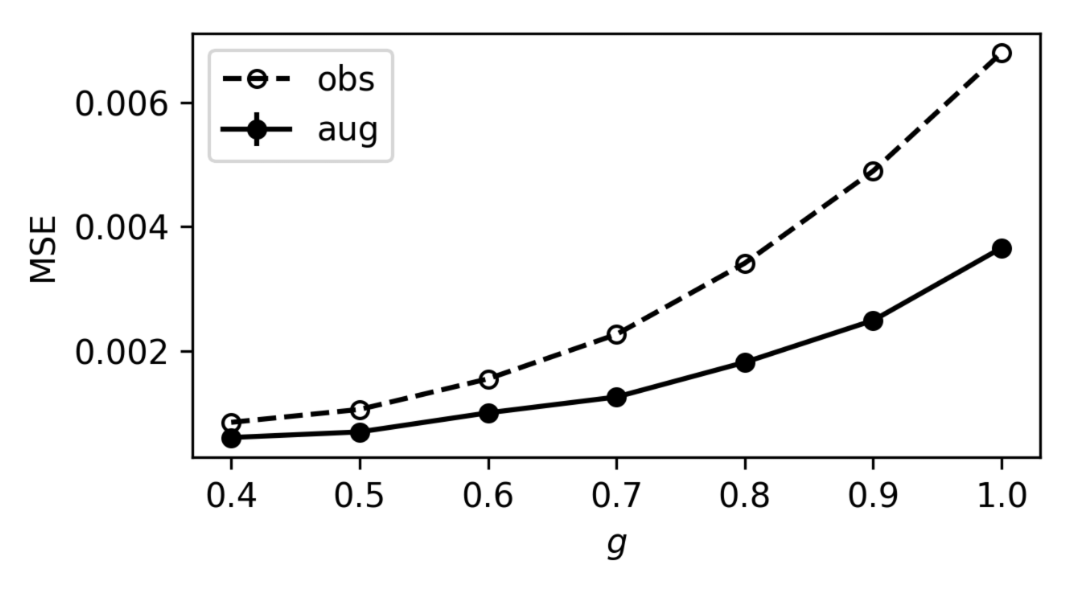}
\caption{\label{g_plot}
Inference with different strengths of bias and interaction parameters. 
The mean square error (MSE) of the inferred parameter values was measured based on the strengths of the bias and interaction parameters, which were chosen from a normal distribution $\mathcal{N}(0, g^{2}/n)$ with system dimension $n$. Inference was performed using only $M=4,000$ observed data (obs; empty circles) and $M^+=100,000$ augmented data (aug; filled circles). The standard deviations of the MSE from 10 ensembles are too small to be visible at the scale of the symbols.}
\end{figure}

\begin{figure}[t]
\centering
\includegraphics[width=0.42\textwidth]{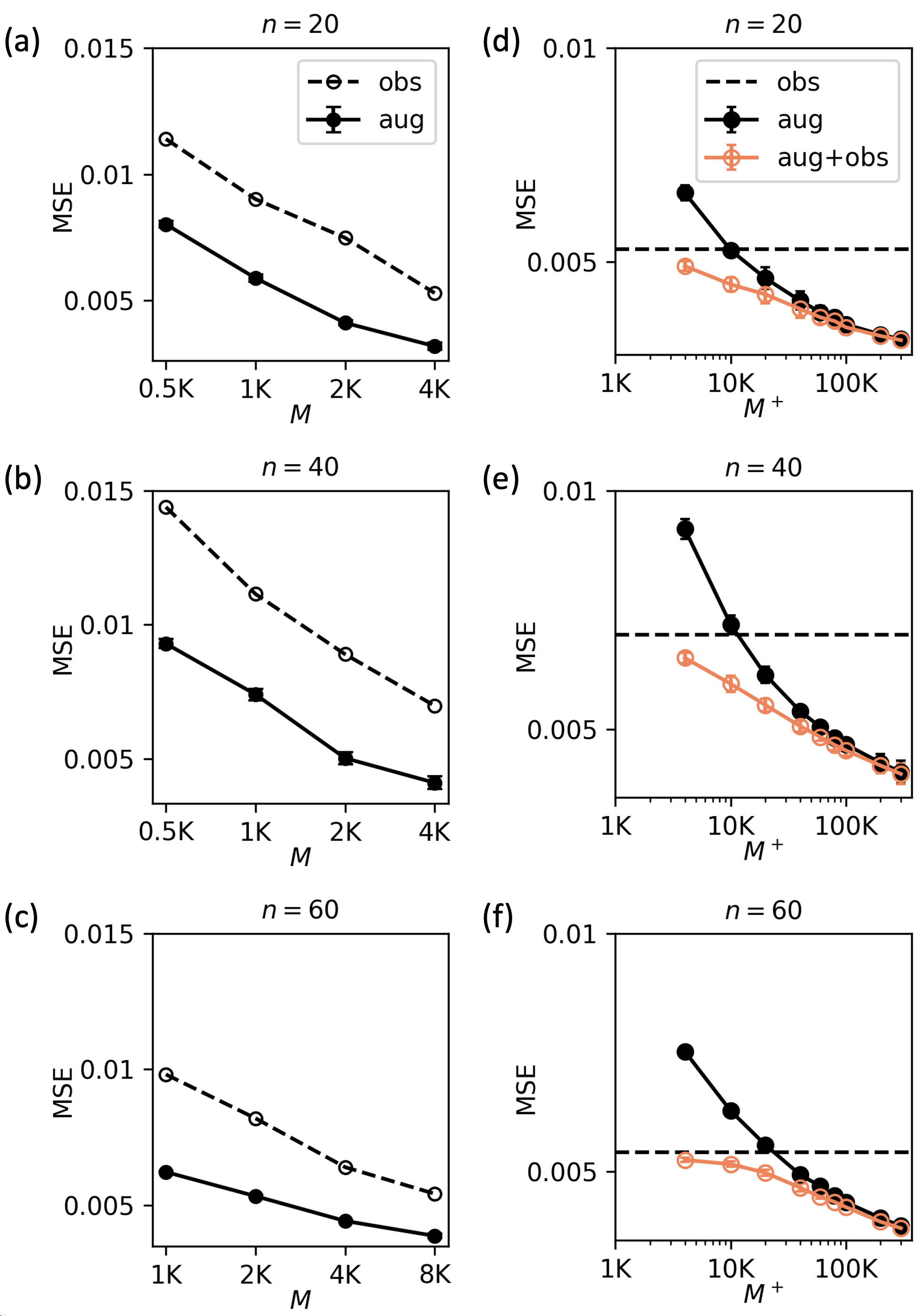}
\caption{\label{data_num}(Color Online)
Inferring performance with the size of observed and augmented data.
(a-c) Mean square error (MSE) of inferred parameter values as a function of the size $M$ of observed data. After training the diffusion model with observed data, the model augmented the data to $M^+=300,000$. The inference error was compared for using only the observed data (empty circles and dashed line) and using the augmented data (filled circles and solid line). All true parameters were chosen to follow a normal distribution $\mathcal{N}(0, g^{2}/n)$ with $g=1$.
(d-f) Inference performance depending on the size $M^+$ of augmented data. The diffusion model for augmenting data was trained with $M=4,000$ observed data (d, e) and $M=8,000$ observed data (f). The horizontal dashed line is the reference MSE of the inference using only the observed data (obs). The inference using only the augmented data (aug; filled black circles) and using both observed and augmented data (aug+obs; empty orange circles) is shown. The dimensions for the Sherrington-Kirkpatrick model are $n=20$ (a, d), $n=40$ (b, e), and $n=60$ (c, f). The standard deviations of the MSE from 10 ensembles are too small to be visible at the scale of the symbols.}
\end{figure}

\section{Results}
\label{results}

We test whether diffusion models can be used for data augmentation in inference problems in physics. First, we validate this idea with a synthetic dataset from the Sherrington-Kirkpatrick model. Second, we apply this approach to real-world data on neuronal activity patterns and demonstrate its effectiveness in reconstructing partially missing neuronal activities.

\subsection{Data augmentation}
Our experimental setup is as follows.
\begin{enumerate}
    \item We initialize the model parameters with random values sampled from Gaussian distributions with zero mean and variance $g^2/n$: $\{b_i, w_{ij}\} \sim \mathcal{N}(0, g^2/n)$, where $g$ controls the magnitudes of the parameters.
    \item Using the energy function $E(\sigma) = -\sum_i b_i \sigma_i - \sum_{i<j} w_{ij}\sigma_i \sigma_j$, we sample $M$ numbers of $\hat{\sigma}$ from the probability distribution, $p(\sigma) = \exp(-E(\sigma))/Z$, using the Markov chain Monte Carlo method.
    After sampling, the parameter values are concealed as the true values $b_i^{\text{true}}$ and $w_{ij}^{\text{true}}$, which will be compared with the inferred values in a later step.
    \item Given the data $\{\hat{\sigma}(\mu)\}_{\mu=1}^M$, we train the diffusion model. Specifically, we use the DDPM approach~\cite{ho2020denoising}. We then generate new samples $\{\hat{\sigma}^+(\mu)\}_{\mu=1}^{M^+}$ from the learned diffusion model. 
    \item Finally, we infer the model parameters using the augmented data with the erasure machine, and verify whether the inferred values are consistent with the true values. 
\end{enumerate}

In the last step, the inference performance is measured by mean square error (MSE):
\begin{equation}
    \text{MSE} = \frac{1}{L} \Bigg(\sum_i (b_i - b_i^{\text{true}})^2 + \sum_{i<j} (w_{ij} - w_{ij}^{\text{true}})^2 \Bigg),
\end{equation}
where $L = n + n(n-1)/2$ is the total number of parameters.
We prepared $M=4,000$ samples of $\hat{\sigma}$ in $n=40$ dimensions. We then trained the neural network $\epsilon_t = \text{NN}(x_t, t)$ for the diffusion model given the original $M$ training data of $\{x_0(\mu)\}_{\mu=1}^M = \{\hat{\sigma}(\mu)\}_{\mu=1}^M$, and generated $M^+ = 100,000$ augmented data $\{\hat{\sigma}^+(\mu)\}_{\mu=1}^{M^+}$. We then examined the inference performance, measured in terms of MSE, as a function of how many epochs the diffusion model has been trained for (see Fig.~\ref{iteration plot}). Initially, as the training progresses, the inference performance improves, as indicated by a decrease in MSE. However, after a certain point, further training leads to a decline in performance, as reflected by an increase in MSE. Finally, the MSE value approaches the inference performance achieved using only the training data.

This issue of overtraining is common in data science. When the diffusion model is insufficiently trained, the quality of the augmented data becomes poor, failing to effectively representing the training data. Conversely, when the model is overtrained, the augmented data may become mere replicas of the training data, which does not improve inference. Thus, finding an appropriate degree of training for the diffusion model is crucial.
One might think this is not a problem because monitoring the MSE can indicate when to stop the training. However, in real-world problems, we do not have access to the true parameter values, making the MSE intractable.

To find a proxy for the MSE, we propose that if the diffusion model appropriately learns the features in the training data, certain statistical properties of the training data and the generated data will be consistent.
\jo{For example, covariance can serve as a potential statistical measure. By comparing the covariance of the training vectors $\sigma$ with that of the generated vectors $\sigma$, we can assess the model's learning efficacy. In the inverse Ising problem, the model defines a scalar energy function for a given vector $\sigma$. For simplicity, we focus on this scalar measure.}
We found that the variance $\text{Var}[E(\sigma)]$ of the energy of each sample $E(\sigma)$ is a good metric for this purpose. Throughout the study, we used the $\{b_i, w_{ij}\}$ inferred solely from the training data, which will be defined later. Given these model parameters, we calculated the energies of the training, test, and augmented samples.

As the diffusion model undergoes more training steps, the variance $\text{Var}[E(\hat{\sigma}^+)]$ decreases.
This is because the inferred parameter values, derived from the training data, get tightly bound to the training data, resulting in more confined energies compared to more unconstrained data
generated by a less trained diffusion model.
However, at a certain point, this trend is flipped and $\text{Var}[E(\hat{\sigma}^+)]$ starts to increase, which may indicate that the diffusion model has been shifted from generalizing to unobserved data to replicating already observed data. Eventually, when the number of epochs goes to infinity, the statistical properties of data generated by the diffusion model, including energy variance, become consistent with those of the training data, as the new data becomes merely a replica of the training data.

To avoid overfitting to the training data, we set aside test data that were not used for training the diffusion model and matched their energy variance to determine the optimal training point.
In the data preparation, we have $\{\hat{\sigma}(\mu)\}_{\mu=1}^{2M}$, with $\{\hat{\sigma}(\mu)\}_{\mu=1}^{M}$ used as training data and $\{\hat{\sigma}(\mu)\}_{\mu=M+1}^{2M}$ used as test data.
This approach helps ensure that the augmented data generalizes well to unseen data.
Indeed, we found that the optimal stopping point, defined by the minimal MSE, can be approximated to a point where the energy variance of the augmented data matches that of the test data (both calculated based on the model parameters inferred from the training data). Note that the former energy variance for the augmented data changes over iterations as the data approaches to the training data, whereas the latter energy variance for the test data is fixed.
A higher quality of the inference was achieved when this energy variance was calculated for the test data and not for the training data.

Adopting this stopping criterion for training the diffusion model, we verified that the augmented data generated by the optimally trained diffusion model enhance the inverse Ising inference (Fig.~\ref{g_plot}). The inference becomes more challenging as the coupling strengths $w_{ij}$ increase with the larger scale parameter $g$. The results show that even for larger $g$, the inference can be improved using $M^+ = 100,000$ augmented data points. It is worth noting that the original dataset contains only $M = 4,000$ samples. 

Now we examine the effects of (i) the size $M$ of the original training data, which affects the learning quality of the diffusion model, and (ii) the size $M^+$ of the augmented data, which impacts the inference performance.
First, we inferred the model parameters using only the training data. As expected, the inference performance improves as more training data is used (Fig.~\ref{data_num}(a-c)). Next, we use a fixed number $M^+ = 300,000$ of augmented data points. Better data augmentation with more training samples can further enhance the inference performance. This result is consistent for small or large systems with $n = 20, 40,$ and $60$.
Second, we fixed the size of the training data at $M=4,000$ for $n=20$ and $40$ systems, and $M=8,000$ for the $n=60$ system.
Compared to the inference performance using only training data, the inference performance improves as more augmented data are used (Fig.~\ref{data_num}(d-f)). One might wonder why not use both the training data and the augmented data together for inference. As expected, doing so further enhances the inference performance.

\begin{figure}[t]
\centering
\includegraphics[width=0.48\textwidth]{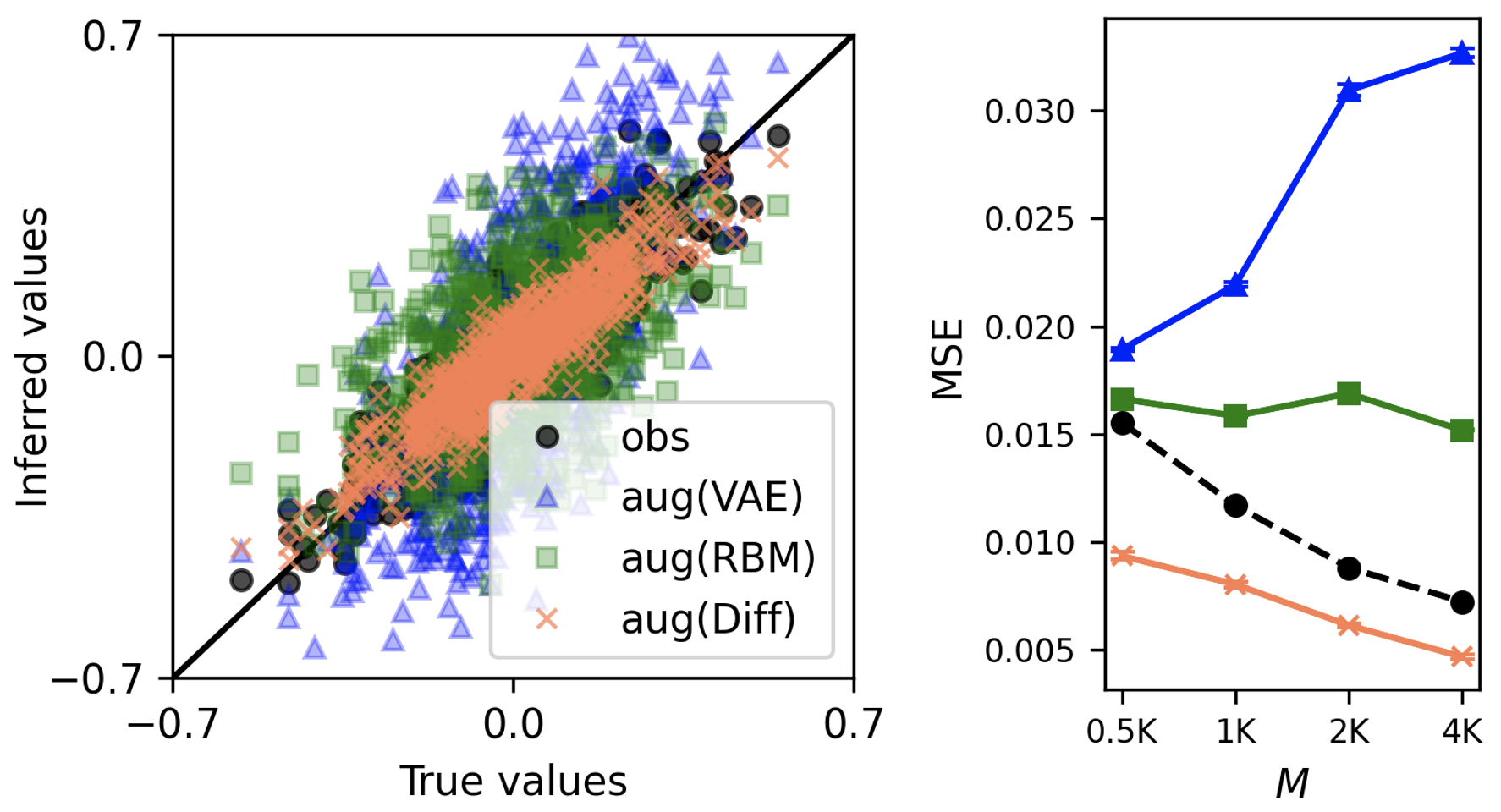}
\caption{\label{comparison}
\jo{(Color Online) Inferring the performance of various generative models. (a) True parameter values are inferred using augmented data from a variational autoencoder (VAE, blue triangles), a restricted Boltzmann machine (RBM, green squares), and a diffusion model (Diff, orange crosses). For comparison, inference results using only the observed data are shown as filled black circles. The generative models are trained with $M = 2,000$ observed data, and parameter inference is performed using $M^+=100,000$ augmented data. (b) The mean square error (MSE) of the inferred parameter values as a function of the size $M$ of observed data. The standard deviations of the MSE from 10 ensembles are too small to be visible at the scale of the symbols. The dimension for the Sherrington-Kirkpatrick model is $n=40$. All true parameters are drawn from a normal distribution $\mathcal{N}(0, g^{2}/n)$ with $g=1$. }}
\end{figure}



\jo{
Next, we evaluate the augmentation performance of the diffusion model in comparison to other generative models. The representative models considered include: (i) an energy-based model, specifically the Restricted Boltzmann Machine (RBM)\cite{hinton2002training}; (ii) a probability-based model, the Variational Autoencoder (VAE)\cite{Kingma2013Dec}; and (iii) a deterministic feedforward neural network, the Generative Adversarial Network (GAN)\cite{goodfellow2014generative}.
Given $M$ observed data, we first train the generative models and generate $M^+$ augmented data. The model parameters are then inferred using this augmented data.
For inverse Ising inference with $n=40$ and $M=2,000$ observed data, the diffusion model outperforms both RBM and VAE (Fig.~\ref{comparison}(a)).
The effectiveness of data augmentation with the diffusion model improves as more observed data is used for training.
In contrast, RBM exhibits relatively consistent performance regardless of the training data size, while VAE performs worse as the amount of training data increases (Fig.~\ref{comparison}(b)).
}

\jo{Note that we exclude GAN results, as its performance is often compromised by model collapse when applied to small-dimensional data, causing it to generate nearly identical samples. Consequently, GAN was not included in our experiments.
VAE tends to produce an averaged representation of the learned data. As the size of the observed dataset increases, this averaging effect becomes more pronounced, leading to a decline in Ising inference performance. This occurs because the model's tendency to generalize and smooth out the data results in less accurate inference of the underlying distribution.}
\jo{The detailed computational implementation of the RBM and VAE is provided in Appendix~\ref{sec:appendix}.}

\begin{figure}[t]
\centering
\includegraphics[width=0.42\textwidth]{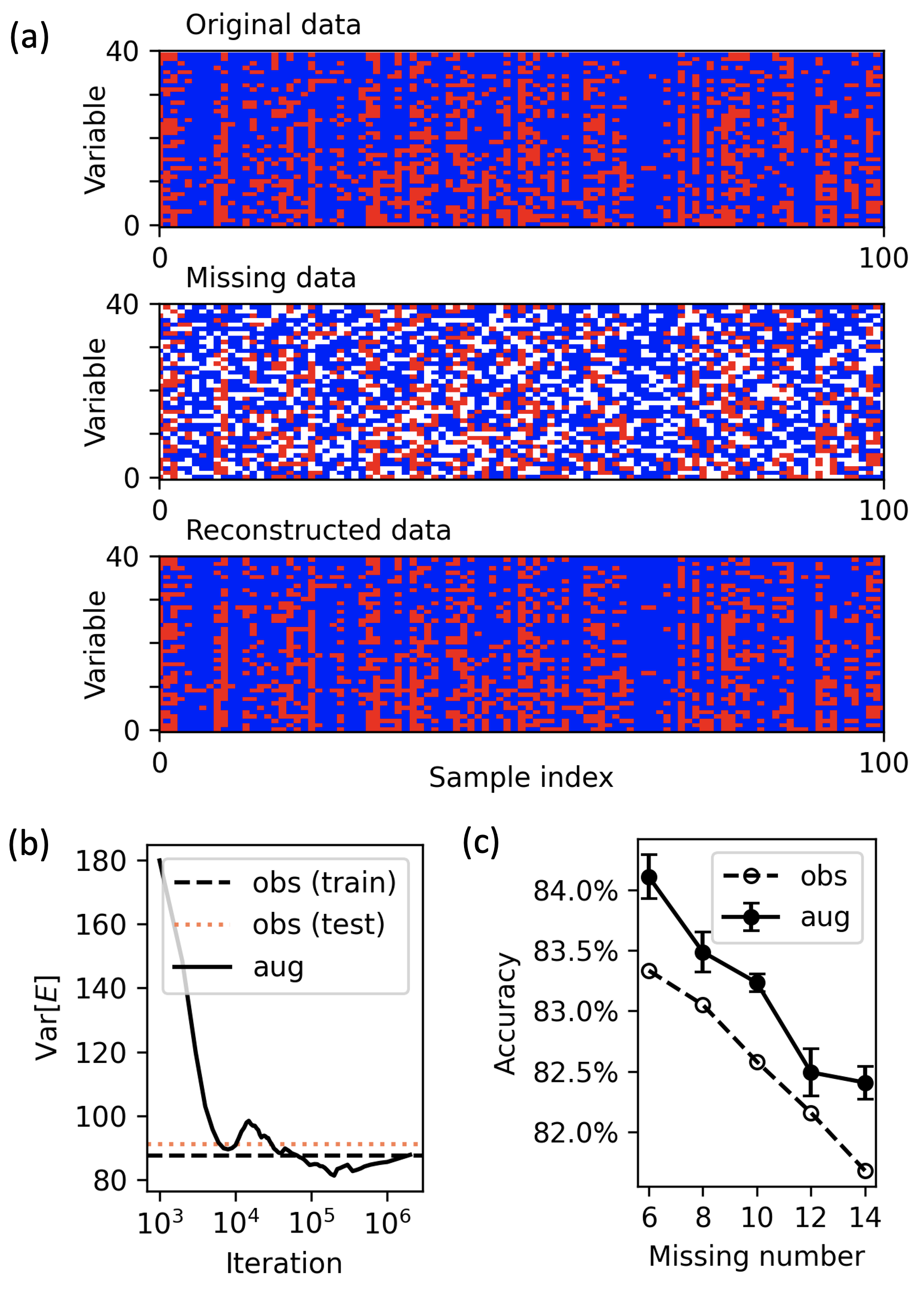}
\caption{\label{neuron experiment}(Color Online)
Reconstruction of neural activities.
(a) Original data consists of binary activity patterns of 40 neurons. One hundred representative patterns are shown, with active states (red squares) and silent states (blue squares). Activities of 14 randomly selected neurons are hidden and represented as missing variables (white squares). The missing data is then reconstructed through inverse Ising inference using a diffusion model to augment data.
(b) The training of the diffusion model is halted once the energy variance of the observed data in the test set (orange dotted line) matches the energy variance of the augmented data (solid line).
(c) Accuracy of the reconstruction depending on the number of missing variables. The accuracy was computed using $M=8,000$ observed data (empty circles) and $M^+=100,000$ augmented data (filled circles). The standard deviation of the accuracy from 10 ensembles is shown, but the standard deviation for the observed data is too small to be visible at the scale of the symbols.}
\end{figure}

\subsection{A real-world problem}

Is the data augmentation actually applicable to real-world problems? To answer this question, we apply the data augmentation to the problem of reconstructing neural activities. 
Temporal neural activities in the tiger salamander (Ambystoma tigrinum) retina have been recorded~\cite{neuron1}.
The dataset consists of neuronal spike trains from 160 neurons of a live salamander stimulated by a film clip of fish swimming, with a bin size of 20 ms (Fig.~\ref{neuron experiment}(a), top). We considered only the 40 most active neurons. 
We model the neural activity at a given time as the Ising model, an equilibrium model that ignores the temporal causal relation between neural activities.
Since the temporal sequence is irrelevant in this model, we randomly arranged the data without considering the order. Initially, we divided the data into 8,000 training samples and 1,000 test samples. We applied the erasure machine to the training samples to infer $\{b_{i}, w_{ij}\}$. Since we do not know the true parameter values in this real-world problem, the inferred local bias $b_i$ and interactions $w_{ij}$ cannot be directly evaluated as good or bad. Therefore, we devised a task where we hide activities of some neurons, referred to as missing activities, and reconstruct these missing activities using the model with the inferred model parameters.

In the test samples, we randomly selected 14 neurons out of the total 40 neurons and designated their activities as missing variables $\hat{\sigma}^m_{k}$ with $k={1, 2, \cdots, 14}$ (Fig.~\ref{neuron experiment}(a), middle). 
The activities of the remaining visible neurons are denoted as observed variables $\hat{\sigma}^o_{i}$ with $i={1, 2, \cdots, 26}$.
The total set of variables can be concisely represented as the combination of missing and observed variables $\hat{\sigma} = (\hat{\sigma}^m, \hat{\sigma}^o)$.
We then reconstruct $\hat{\sigma}^m$ to maximize the probability $p(\hat{\sigma}^{m}|\hat{\sigma}^{o}) \propto p(\hat{\sigma}^{m},\hat{\sigma}^{o})$ (Fig.~\ref{neuron experiment}(a), bottom). The accuracy of matching the missing variables was measured to evaluate the reconstruction performance.
In the inverse Ising inference, we can use only the training data ($M=8,000$) or use augmented data ($M^+=100,000)$) generated by the diffusion model.
During the training of the diffusion model, we monitor the energy variance and stop the training when the energy variance of the augmented data matches that of the test data, as previously defined (Fig.~\ref{neuron experiment}(b)).
Finally, we verified that the augmented data improved the reconstruction of the missing values (Fig.~\ref{neuron experiment}(c)). The reconstruction accuracy diminishes as the number of missing variables increases, but the augmented data consistently improves reconstruction accuracy regardless of the number of missing variables.

\section{Discussion}
\label{discussion}

Machine learning can be viewed as a framework for probability modeling of data spaces, typically represented using neural networks. The ``score" function plays a pivotal role in this framework, capturing the gradient of probability distributions with respect to small perturbations in configurations, thus representing the scalar probability as vector flows. This approach has proven exceptionally effective in modeling underlying probability distributions and generating new samples that adhere to these distributions.
\jo{This gradient-based modeling proves advantageous for data representation compared to other generative models. Energy-based models suffer from the computational burden of estimating the partition function, while probability-based variational models and deterministic feedforward models struggle to generate diverse samples. Variational autoencoders often produce overly smoothed, averaged data, whereas generative adversarial networks tend to generate a limited set of samples, leading to mode collapse.}

In this study, we explore whether the remarkable generative capabilities of diffusion models can be leveraged to address challenges in physics. We demonstrate their potential by applying these models to augment data and enhance inference tasks. Specifically, we show improvements in inverse Ising inference through data augmentation and apply these techniques to neural activity modeling, including the reconstruction of missing values in activity data. Our findings suggest that diffusion models offer valuable tools for advancing both theoretical and applied physics problems, highlighting their broad applicability and effectiveness.

While our study focused on binary data, the approach can be extended to accommodate multi-state and even continuous data. Notably, data augmentation using diffusion models can be effectively adapted to generate multiple sequence alignment (MSA) data, which is critical for protein folding problems. This task closely resembles the inverse Ising inference, differing primarily in its multi-state nature beyond binary~\cite{haldane2019influence}. Since MSA is inherently linked to corresponding three-dimensional molecular structures, mutations in the sequence must reflect site-specific correlations~\cite{morcos2011direct}. Conversely, this mutation information encodes insights into molecular structures. Recently, MSA datasets have significantly improved protein structure predictions in tools like AlphaFold~\cite{roney2022state}. However, the availability of sufficient MSA data remains a challenge, making the augmentation of MSA datasets an essential task. 
This extension highlights the versatility of diffusion models in addressing complex data challenges across diverse domains.
Furthermore, the inverse Ising inference discussed in this study focuses on equilibrium spin states. However, it can be extended to account for the temporal dynamics of spins~\cite{Zeng2011Apr,Zeng2013May}. Future work could explore leveraging diffusion models to augment time-series data, opening new possibilities for dynamic systems analysis.

Finally, we note that this study relies on diffusion models designed for continuous variables. To generate binary data, we employed a thresholding technique during the final stage of sample generation, known as the analog bit approach~\cite{chen2023analog}. Since this method proved effective for the inverse Ising inference, we adopted it in our work. 
However, recent advancements in discrete diffusion models have introduced promising alternatives for generating discrete data. One class of these models employs a Markov jump process as the diffusion mechanism for discrete variables~\cite{park2024textit}, while another class updates continuous parameters of probability models to represent discrete variables~\cite{graves2023bayesian}. 
Given these developments, it is natural to anticipate the application of true discrete diffusion models for augmenting discrete data. Exploring this direction remains an exciting avenue for future research.

\begin{acknowledgments}
We thank Yong-Hyun Park for helpful discussions and comments.
This work was supported by the Creative-Pioneering Researchers Program through Seoul National University, and the NRF Grant No. 2022R1A2C1006871 (J.J.). \\
\end{acknowledgments}

\appendix
\section{Model details}
\label{sec:appendix}

\jo{{\it Diffusion model.}
We implement $\epsilon_t = \text{NN}(x_t, t)$ using a multilayer perceptron architecture, inspired by U-Nets.
Specifically, for an input of dimension $n=40$, the network consists of an encoder with three layers, having node sizes of 128, 256, and 512, followed by a decoder with four  layers, with node sizes of 512, 512+512, 256+256, 128+128. The additional nodes, denoted with plus in the decoder, indicate residual connections from the encoder, which, based on our empirical observations, are crucial for enhancing performance. 
For a smaller input dimension of $n=20$, we achieve good performance even without residual connections.
For a larger input dimension of $n=60$, we retain the same network architecture but increase the nose sizes as follows: 128 $\to$ 200, 256 $\to$ 400, and 512 $\to$ 600.
The activation function used is Gaussian Error Linear Units (GELU). The model is trained with a linear schedule~\cite{ho2020denoising}, defined as $1-\alpha_t = 0.0001+0.02t/T$ and employs sinusoidal position embeddings for encoding time $t$ in $\epsilon_t = \text{NN}(x_t, t)$. Optimization is performed using the Adam optimizer with a learning rate of 0.0001.}

\jo{{\it Restricted Boltzmann machine.} For an input dimension of $n=40$, we used 20 hidden nodes and trained the model using contrastive divergence with one step (CD$_1$)~\cite{hinton2002training}. Momentum optimization was applied with a learning rate of 0.05, a momentum coefficient of 0.5, and a weight decay of 0.0001 to mitigate overfitting.}

\jo{{\it Variational autoencoder.} For an input dimension of $n=40$, we used a network architecture of 40-30-20-10-20-30-40. Then, the model was optimized using the Adam optimizer with a learning rate of 0.001.}

\bibliographystyle{apsrev4-1}
\bibliography{DBB}

\end{document}